# Transparent hardware synthesis of Java for predictable large-scale distributed systems

Extended Abstract

Ian Gray, Yu Chan, Jamie Garside, Neil Audsley, Andy Wellings
Real-Time Systems Group, Department of Computer Science, University of York
{ian.gray, yc522, jamie.garside,
neil.audsley, andy.wellings}@york.ac.uk

*Abstract*—The JUNIPER project is developing a framework for the construction of large-scale distributed systems in which execution time bounds can be guaranteed. Part of this work involves the automatic implementation of input Java code on FPGAs, both for speed and predictability. An important focus of this work is to make the use of FPGAs transparent though runtime co-design and partial reconfiguration. Initial results show that the use of Java does not hamper hardware generation, and provides tight execution time estimates. This paper describes an overview the approach taken, and presents some preliminary results that demonstrate the promise in the technique.

## I. Introduction

*Big Data* is the term used for application requirements that cannot be met using existing data processing techniques, because of either the sheer scale of the input data, or the timing requirements that are placed on the system. As a result, FPGAs are starting to be deployed into data centres to exploit the large parallelism and low latency that they can offer. However, effective use of FPGAs requires significant specialist knowledge; of hardware description languages (HDLs), complex vendor tools, and high-level synthesis (HLS) systems.

As a response to this, the JUNIPER project is developing a framework for *soft real-time* Big Data systems that includes technology for automatic translation of user Java code to FPGA hardware. Rather than simply "fast enough", JUNIPER views real-time to mean that the correctness of the data is dependent on both its value and the time by which it is delivered. Hardware translation is used because hardware components have tighter bounds on their worst case response time, and are very useful for the construction of more predictable systems. Also FPGA implementations tend to display greater performance than their Java equivalents. Unlike systems that focus on using high-level synthesis to create a highly-optimised hardware implementation of a key system component, the key contribution of the JUNIPER approach is that it aims to allow totally transparent FPGA acceleration through the use of online configuration and partial dynamic reconfiguration.

The input language to the JUNIPER system is either standard Java 8, or Java written with the Real-Time Specification for Java (RTSJ). The use of Java is motivated by its common use in the large-scale data processing domain. Systems such as Hadoop are written in Java, and Spark and Storm are written partially in Java, and are implemented on the Java Virtual Machine (JVM). JUNIPER is also compatible with other JVM languages, such as Clojure and Scala.

## II. Programming model

The JUNIPER API is a Java 8 API for supporting large-scale computing environments, such as clusters ("cloud computing") and high-performance computers. The full details of the JUNIPER model are outside of the scope of this extended abstract and are detailed in existing work [1], [2]. In brief, JUNIPER allows the programmer to split their code into units which may be deployed into separate compute nodes. Inter-node communications, data flow, and storage, are automatically handled by the API layer.

In addition to this, JUNIPER programs can use a concept called *Locales*. Rather than placing threads and data manually using affinities, a locale is a software-level element which is used to inform the JVM and platform that the threads and data inside a locale will be tightly-coupled and so should be located as closely together as possible. These bundled threads and data items are then dynamically mapped to subsets of the target architecture, and for the purpose of this work may also be deployed to FPGAs. Online FPGA compilation and partial reconfiguration allows the system to search for a suitable mapping. This helps to solve a common problem with general-purpose acceleration of a high-level language in which it can be difficult to determine the parts of the application that should be accelerated for the largest gain.

## III. Implementation strategy

The JUNIPER toolflow is shown in figure 1. The input Java (or other JVM language) is translated to C for native compilation by a real-time JVM called JamaicaVM [3]. This approach supports both standard Java and real-time Java, and allows for more predictable real-time behaviour (including real-time garbage collection). A tool called *caicos* then manages the creation of complementary hardware (FPGA) and software (host) projects. On the hardware side, the high-level synthesis tool Vivado HLS is used to translate from C to HDL.

A key requirement of this work is that use of the FPGA must be transparent to the programmer. Before the translated C can be passed to Vivado HLS, significant rewriting must be performed in order to ensure that efficient hardware is produced. First, all global memory accesses (the Java heap) from

This work has received funding from the European Union's Seventh Framework Programme under grant agreement FP7-ICT-611731

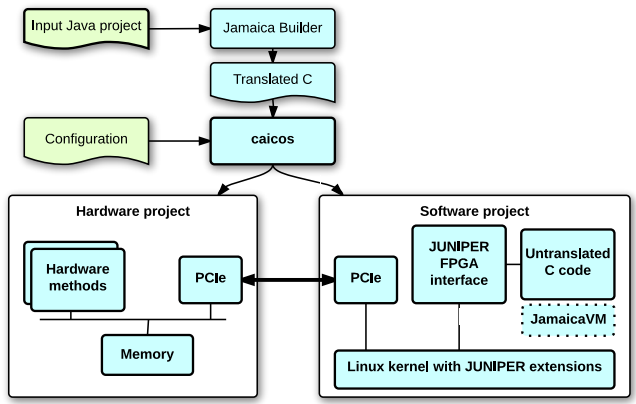

Fig. 1. The hardware and software flows in JUNIPER.

the translated C must be rewritten into AXI bus transactions. The use of pointers is avoided to ensure efficient synthesis. Secondly, because abstract or interface method calls may dynamically dispatch to different implementations based on the type of the called object, JamaicaVM and caicos perform static analysis to determine exactly which subset of methods may be called to minimise multiplexer use. Finally, untranslatable software (VM calls, native methods, etc.) are translated into a 'system call' in which the hardware calls back to the host processor over the PCIe bus to execute the required function.

The only limitation on input software is that exceptions are currently not supported inside translated methods because of the hardware complexity they introduce. It is possible to reduce this through static analysis, but this remains further work.

## IV. Dynamic Acceleration

Due to space constraints on the FPGA, most of the time it will not be possible to offload all code to the FPGA simultaneously. Whilst JUNIPER allows the designer to pick a fixed subset for hardware implementation, it is also developing a dynamic acceleration approach to make the acceleration transparent to the developer through the use of online compilation, synthesis, and partial reconfiguration.

In the target domain of commercial large scale data systems, applications tend to be permanently running and can afford to dedicate a compute node to performing speculative synthesis and implementation. JUNIPER uses this to explore the design space automatically, and uses dynamic reconfiguration to swap new test bitfiles in to the running application. This is facilitated by extensive online monitoring that is provided by the JUNIPER framework. Once an improved design is found, the system will update and redeploy itself, perhaps onto fewer computer nodes if it can still guarantee its required response times.

## V. Preliminary results

As this represents work in progress only relatively small filters and methods have been tested, however some interesting preliminary results have already been discovered. Table I shows comparisons between hand-written C and the JUNIPER approach (on a Xilinx Virtex 7 series device). It can be seen that the use of Java generally only imposes a small logic area

TABLE I. Comparison of hand-developed C and JUNIPER (naïve synthesis, without manual optimisation)

| Function | Hand-developed C + HLS | | Java + JamaicaVM + HLS | |
|---|---|---|---|---|
| | LUTs | Latency | LUTs | Latency |
| Vector sum | 113 | 507 | 175 | 511 |
| Collatz evaluation | 293 | 278 | 383 | 282 |
| MD5 hash | 1675 | 3463 | 272 | 676 |
| FIR filter | 298 | 183 | 283 | 121 |

and latency overhead (due to additional bus logic and memory access routines).

The table also shows one benefit of the approach. The MD5 result shows a huge improvement in both speed and area from using Java over C. This is because all of these numbers are before any hand-optimisation of synthesis directives. In the case of MD5, manual unrolling and function inlining can reduce the hand-developed C version to be similar in size and speed to the JamaicaVM version, but this requires specialist knowledge and is not transparent to the user. A lot of the overhead is in the C version's use of pointers, something which is removed by the restricted stack-machine of Java bytecode.

In all of these results we can see that the generated hardware has a specific latency value, rather than a range. With fixed inputs we can be certain down to the clock cycle about how long a piece of hardware will take to execute. Uncertainty can be introduced through memory latency or bus/network latency as with software implementations. These results show that there is potential for the JUNIPER acceleration approach. Evaluation of large-scale applications is currently being undertaken.

## VI. Conclusion

The JUNIPER platform is an approach to building the next generation of Big Data systems which can provide design-time guarantees about their response times and performance metrics. To do this, the platform includes a range of real-time technologies, including transparent integration of FPGAs for speed and predictability.

Initial results show that the use of Java to accelerate software does not add significant overheads, and in fact when code becomes more complex and 'C-like' the JUNIPER toolflow can give better results unless manual expertise is then applied. It also provides tighter execution time estimates. This paper describes the work currently under way, the approach being developed, and presents some preliminary results that demonstrate the promise in the technique.